\documentclass[onecolumn,showpacs,floatfix,aps,nofootinbib]{revtex4}
\usepackage[dvips]{graphics,color}
\usepackage{epsfig}\usepackage{float}
\RequirePackage{amssymb}
\usepackage{makeidx}
\usepackage[brazil]{babel}
\usepackage[latin1]{inputenc}
\usepackage{graphicx}
\usepackage{indentfirst}
\usepackage{color}
\newcommand{\be}{\begin{equation}}
\newcommand{\ee}{\end{equation}}

\newcommand{\ba}{\begin{eqnarray}}
\newcommand{\ea}{\end{eqnarray}}

\begin{document}

\title{Towards an hybrid compactification with a scalar-tensor global cosmic string}

\author{M. C. B. Abdalla$^{1}$}
\email{mabdalla@ift.unesp.br}
\author{M. E. X. Guimar\~aes$^{2}$}
\email{emilia@if.uff.br}
\author{J. M. Hoff da Silva$^{1}$}
\email{hoff@ift.unesp.br}

\affiliation{1. Instituto de F\'{\i}sica Te\'orica, Universidade
Estadual Paulista, Rua Pamplona 145 01405-900 S\~ao Paulo, SP,
Brazil}

\affiliation{2. Instituto de F\'{\i}sica, Universidade Federal
Fluminense, Niter\'oi-RJ, Brazil}


\begin{abstract}
\vspace{2cm}
\begin{center}
{\it Abstract}
\end{center}

We derive a solution of the gravitational equations which leads to
a braneworld scenario in six dimensions using a global cosmic
string solution in a low energy effective string theory framework.
The final spacetime is composed by one warped brane with
$\mathbb{R}^{(3,1)}\times S^{1}$ topology and a power law warp
factor, and one noncompact extra dimension transverse to the
brane. By looking at the current experimental bounds, we find a
range of parameters in which, if the on-brane dimension has an
acceptable size, it does not solve the hierarchy problem. In
another example this problem is smoothed by the Brans-Dicke
parameter.
\end{abstract}
\maketitle

\newpage
\section{Introduction}

In the recent years the amount of works dealing with the
possibility of large extra dimensions in the universe was crescent
\cite{first,second}. The general characteristic of these type of
models is to treat our universe as a four dimensional brane (a
3-brane) in a higher dimensional spacetime \cite{RS}. In the
Randall-Sundrum model \cite{RSII}, a $5-D$ realization of the
Horava-Witten solution \cite{HW}, the hierarchy problem can be
solved by introducing an appropriated exponential warp factor in
the metric. The general structure of the Randall-Sundrum model is
given by two mirrors domain walls (the branes) embedded into an
orbifold in the extra noncompact directions. Following the
approach of the use of topological defects as the generator of
bulk-brane structure, an alternative scenario was proposed in
\cite{RUTH} in which a $6-D$ spacetime comes from a global cosmic
string in the Einstein gravity. These type of models are
characterized by the global long range signature of the
gravitational field, which restricts the compactification
possibility of extra dimensions.

The program of compactification in General Relativity was extended
to more dimensions using global defects \cite{VILE}. In the
reference \cite{CEG}, for instance, a higher codimension case is
analyzed where the hierarchy is introduced by hand, with a power
law warp factor. Besides, the scenario presents an important
characteristic revealed by the shape of the spacetime: far from
the brane, the geometry is almost unaffected by the presence of
the brane. On the other hand the backreaction effects are
important near the brane. In this case, a naked singularity is
expected in the zero width limit.

Frequently, in the braneworld framework it is necessary to
introduce scalar fields in the bulk in order to localize the
branes \cite{B6D,Ram/Wands}. Apart of this, advances in string
theory point into a presence of a dilatonic scalar field in the
low energy spectrum of recovered gravity \cite{cordeiros}. In this
context it seems natural to work in theories such as the
Brans-Dicke one \cite{BD}, which is the easiest scalar-tensor
gravity theory, having essentially a scalar field intermediating
the gravitational force together with the usual purely tensor
field $g_{\mu\nu}$. In such framework, a braneworld model using a
local cosmic string was proposed in \cite{a gente}, however due to
the complicated functional form of the field equations, the main
argumentation was based on dynamical systems tools.

The main purpose of the present paper is to extend ideias of the
ref. \cite{a gente} and to present a solution of the
Einstein-Brans-Dicke (EBD) equation which points into an exotic
(and hybrid) compactification. Here, we use a global cosmic string
model coupled to  gravity. Since the establishment of a braneworld
model is, at least initially, mainly a classical issue, we focus
our arguments in the gravitational aspects of the Brans-Dicke
theory. We start with a $(p+3)$-dimensional set up, looking for a
solution of the EBD system with the energy-momentum of a global
cosmic string {\it outside} of the string core. We do not consider
the mechanism of string formation. Instead, we just establish the
spacetime outside the string core. As usual for this type of
defect, the bulk-brane structure obtained here has six dimensions.
This type of six-dimensional scenario was previously studied in
General Relativity with an explicit local vortex source
\cite{GMS}. Besides, a blown-up brane was analyzed in
\cite{UE,UEII}, where a $\mathbb{Z}_{2}$ symmetry connects the
solutions inside and outside the brane. Returning to the present
work, after solving the EBD equation we restrict the range of
integrations constants in order to study some specific examples.

The final scenario appears to be strongly dependent on the
relationship between such constants and our choice is made in
order to analyze a concrete example. If, by one side, a concrete
case provides a more insightful situation from the physical point
of view, from another side we pay the price of  obtaining
conclusions which are restricted to the case in question.
Nevertheless, we should keep the study of  specific examples,
since in this way we can perform a more critical analysis. The
line element in such examples shows, then, an hybrid
compactification with one 4-brane with topology given by
$\mathbb{R}^{(3,1)}\times S^{1}$ and a power law warp factor. The
spacetime far from the brane is not well-behaved, as well as in
the limit of the transverse dimension going to the brane. We
remark, however, that the solution is obtained for $r\geq r_{b}$,
where $r$ is codimension and $r_{b}(>0)$ is the string radio (the
brane width). This last characteristic is, in some sense, similar
to the case found in the reference \cite{CEG} for higher
codimension braneworld in General Relativity.

The final result also shows an important relation between the
radius of the extra on brane dimension and the warp factor: in one
example, as the warp factor decreases with the transverse
dimension, the extra dimension size grows, and vice-versa. Such a
behavior is not new in Brans-Dicke braneworld models \cite{B6D}.
In another example, this opposite behavior remains but it is
slightly modified. In reference \cite{a gente}, it was clear that
the Brans-Dicke scalar field opens new possibilities of adjustment
in the solution of the hierarchy problem via the warp factor.
Nevertheless, by looking at the current lower bound of the
Brans-Dicke parameter \cite{exper}, we find a range of parameters
in which the extra on-brane dimension has an acceptable size, but
it does not solve the hierarchy problem in this (pre)model. Such a
characteristic is also present in other models \cite{UE}. In the
least example analyzed here it appears a more interesting warp
factor, in which the hierarchy problem is smoothed by the
Brans-Dicke parameter.

The paper is organized as follows. First, we generalize the
solution for a global string in four dimensions obtained by
Boisseau and Linet in \cite{Linet} for six dimensions, starting
from $(p+3)$-dimensions and after setting $p=3$ in order to
recognize a bulk-brane set up. Then we work in the sense of
constrain the range of parameters in order to study some concrete
examples in this pre-model. Finally, in the last section we
conclude summarizing our main results and discussing the (still)
open issues to be considered in the future, as the generalization
of the solution to $r\leq r_{b}$ and the moduli stabilization,
both highly necessary in a realistic model.

\section{Towards the Model}

We shall first analyze a global cosmic string model in
$(p+3)$-dimensions within the Brans-Dicke gravity (hereafter
called simply as BD). Then, we fix the dimension in order to
recognize the bulk-brane structure. It is more convenient to work
with the non-physical metric, i.e.,  in the Einstein's frame,
where the tensorial field $g_{\mu\nu}$  decouples from the scalar
field (the ``dilaton") $\phi$, related to the physical one by
$\tilde{g}_{\mu\nu}=e^{2\alpha \phi}g_{\mu\nu}$. By definition,
the physical BD scalar field is
$\tilde{\phi}=\frac{1}{\mathcal{G}}e^{-2\alpha \phi}$, where
$\alpha$ is a dimensionless parameter related to the BD one, $w$,
by $\alpha^{2}=\frac{1}{2w+3}$ and $\mathcal{G}$ is correlated to
the Planck mass in $(p+3)$ dimensions.

The ansatz for the non-physical metric will respect the symmetry
produced by a straight $U(1)$ global cosmic string, i. e., a
cylindrical symmetry \ba
ds^{2}=-g(r)dt^{2}+dr^{2}+g_{1}(r)d\theta^{2}+g(r)dz_{i}^{2}\label{1},\ea
where $i=1...p$. Besides, outside the string core radius, the
stress tensor is given by \cite{Linet}
\be{T^{r}_{r}=T^{z}_{z}=T^{t}_{t}=-T^{\theta}_{\theta}=-\sigma(r)},\label{2}\ee
where $\sigma$ is a strictly positive function to be determined.
Note that with such settings we are considering the brane at the
origin and the solution will be valid for $r\geq r_{b}$, where
$r_{b}$ is the string radius core (the brane width).

In terms of $g_{\mu\nu}$ the BD equation is \ba
R_{\mu\nu}-\frac{1}{2}g_{\mu\nu}R=2\partial_{\mu}\phi\partial_{\nu}\phi+g_{\mu\nu}g^{\gamma\delta}
\partial_{\gamma}\phi\partial_{\delta}\phi
+8\pi\mathcal{G}T_{\mu\nu},\label{3} \ea or, in a more convenient
way,
\be{R^{\mu}_{\nu}=2\partial^{\mu}\phi\partial_{\nu}\phi+8\pi\mathcal{G}
\Bigg(T^{\mu}_{\nu}-\frac{\delta^{\mu}_{\nu}T}{p+1}\Bigg)}\label{4}
\ee and the scalar BD field satisfies the wave equation \be{\Box
\phi=-4\pi\mathcal{G}\alpha T}\label{box}.\ee The stress tensor is
no longer conserved in this frame. In fact,
\be{\nabla_{\mu}T^{\mu}_{\nu}=\alpha
T\partial_{\nu}\phi},\label{5}\ee which is equivalent to
\be{\frac{\sigma'}{\sigma}+\frac{g_{1}'}{g_{1}}+\frac{(p-1)}{2}\frac{g'}{g}=2\alpha
\phi'}.\label{6}\ee Its solution is given by
\be{\sigma=\frac{\sigma_{0}}{8\pi\mathcal{G}}\frac{e^{2\alpha\phi}}{g_{1}
g^{\frac{(p-1)}{2}}}}\label{7},\ee where $\sigma_{0}$ is an
arbitrary positive constant. From equation (\ref{box}), and using
(\ref{7}), one has \be{
\phi''+\frac{1}{2}\Bigg[\frac{g'(p+1)}{g}+\frac{g_{1}'}{g_{1}}
\bigg]\phi'=\frac{\alpha
\sigma_{0}e^{2\alpha\phi}}{g_{1}g^{\frac{(p-1)}{2}}}.\label{8}}\ee

The set of equations (\ref{4}), with the metric (\ref{1}), gives
\ba
\frac{-g''}{2g}+\frac{(1-p)g'^{2}}{4g^{2}}-\frac{g'g_{1}'}{4gg_{1}}=8\pi\mathcal{G}\sigma
\Bigg(\frac{1-p}{1+p}\Bigg)\label{9}, \ea \ba
\frac{-g_{1}''}{2g_{1}}+\frac{g_{1}'^{2}}{4g_{1}}-\frac{(p+1)g_{1}'g'}{4g_{1}g}=8\pi\mathcal{G}\sigma
\Bigg(\frac{3+p}{1+p}\Bigg)\label{10}, \ea \ba
-\frac{(p+1)}{2}\frac{g''}{g}+\frac{(p+1)}{4}\frac{g'^{2}}{g^{2}}-\frac{g_{1}''}{2g_{1}}+\frac{g_{1}'^{2}}{4g_{1}^{2}}=8\pi\mathcal{G}\sigma
\Bigg(\frac{1-p}{1+p}\Bigg)+2\phi'^{2}\label{11}. \ea Now,
defining $\overline{u}^{2}=g_{1}g^{(p+1)}$, one can rewrite the
equations (\ref{8})-(\ref{10}), respectively, as
\be{\frac{d}{dr}\Big(\overline{u}\frac{d\phi}{dr}\Big)=\alpha
\sigma_{0}\overline{u}\frac{e^{2\alpha\phi}}{g_{1}g^{\frac{(p-1)}{2}}}
},\label{12}\ee \be{\frac{d}{dr}\Big(\frac{\overline{u}}{g}\frac{d
g}{dr}\Big)=-16\pi\overline{u}\sigma\Bigg(\frac{1-p}{1+p}\Bigg)
},\label{13}\ee
\be{\frac{d}{dr}\Big(\frac{\overline{u}}{g_{1}}\frac{d
g_{1}}{dr}\Big)=-16\pi\overline{u}\sigma\Bigg(\frac{3+p}{1+p}\Bigg)
}.\label{14}\ee Substituting the equations (\ref{9}) and
(\ref{10}) into (\ref{11}) we have \ba
\frac{p(p+1)}{4}\frac{g'^{2}}{g^{2}}+\frac{(p+1)}{2}\frac{g_{1}'g'}{g_{1}g}=8\pi\mathcal{G}\sigma(p-3)+2\phi'^{2}\label{15}.
\ea

It is useful to introduce a new radial coordinate, $\bar{r}$, just
like in \cite{Linet} by defining
$\overline{u}\frac{d{\bar{r}}}{dr}=1$. The new coordinate runs in
the domain $-\infty <\bar{r}<+\infty$, and in terms of this new
variable the metric becomes (suppressing the bar label)\ba
ds^{2}=-g(r)dt^{2}+g^{(p+1)}g_{1}dr^{2}+g_{1}(r)d\theta^{2}+
g(r)dz_{i}^{2} \label{16}.\ea  Note that in this new coordinate
system the brane is located at $r\rightarrow -\infty$ and the
solutions will be valid in the region $r\geq -r_{b}$. Rewriting
the equations (\ref{12})-(\ref{14}), we have
\be{\frac{d^{2}\phi}{dr^{2}}=\alpha
\sigma_{0}g^{\frac{(p+3)}{2}}e^{2\alpha\phi}}\label{17},\ee
\be{\frac{d}{dr}\Big(\frac{1}{g}\frac{d
g}{dr}\Big)=\frac{-2(1-p)\sigma_{0}g^{\frac{(p+3)}{2}}e^{2\alpha\phi}}{(1+p)}},\label{18}\ee
\be{\frac{d}{dr}\Big(\frac{1}{g_{1}}\frac{d
g_{1}}{dr}\Big)=\frac{-2(p+3)\sigma_{0}g^{\frac{(p+3)}{2}}e^{2\alpha\phi}}{(1+p)}},\label{19}\ee
while the equation $(\ref{15})$ remains the same, apart of a
$\bar{u}^{2}$ term multiplying the first part of the right-hand
side.

From the equations (\ref{17}) and (\ref{19}) we obtain
\be{g_{1}=g_{1}^{0}e^{\kappa_{1}r}e^{\frac{-2}{\alpha}\big(\frac{3+p}{1+p}\big)\phi}}\label{20}.\ee
Equations (\ref{17}) and (\ref{18}) together show that
\be{g=g^{0}e^{\kappa
r}e^{\frac{-2}{\alpha}\big(\frac{1-p}{1+p}\big)\phi}}\label{21},\ee
with $\kappa_{1}$ and $\kappa$ being arbitrary constants while
$g_{1}^{0}$ and $g^{0}$ are positive constants. We should
emphasize that, if $p=1$, the expressions for $g$ and $g_{1}$
agree with the solutions found in \cite{Linet}. Looking at the
expression (\ref{16}), it is clear that if $p=3$ we arrive into a
braneworld scenario. From now on, we shall fix $p=3$, i.e., we are
dealing with a six dimensional model. Substituting the expressions
obtained for $g$ and $g_{1}$ into equation (\ref{15}) one has \ba
\Bigg(2+\frac{3}{\alpha^{2}}\Bigg)\phi'^{2}-\frac{2\kappa_{1}}{\alpha}
\phi' - \kappa\Bigg(3\kappa+2\kappa_{1}\Bigg)=0 \label{22},\ea
which  solution is given by $\phi =\phi_{0}r+\phi_{1}$, where
$\phi_{1}$ is an arbitrary constant (hereafter $\phi_{1}=0$
without loose of generality) and \ba
\phi_{0}=\frac{\kappa_{1}}{\alpha
\big(2+\frac{3}{\alpha^{2}}\big)}\pm
\frac{1}{2\big(2+\frac{3}{\alpha^{2}}\big)}\Bigg[\frac{4\kappa_{1}^{2}}{\alpha^{2}}
+4\kappa\big(3\kappa+
2\kappa_{1}\big)\big(2+\frac{3}{\alpha^{2}}\big)
\Bigg]^{1/2}.\label{u1}\ea

With $\phi$ one can determine exactly the behavior of $g_{1}$ and
$g$ functions and then we can write down the expression for the
metric. The physical line element for the spacetime is \ba
\tilde{ds^{2}}&=&\left.
g^{0}e^{\big[\kappa+\big(\frac{1+2\alpha^{2}}{\alpha}
\big)\phi_{0}\big]r}\eta_{\mu\nu}dx^{\mu}dx^{\nu}+g_{1}^{0}e^{\big[\kappa_{1}+\big(\frac{-3+2\alpha^{2}}{\alpha}
\big)\phi_{0}\big]r}d\theta^{2}\right.\nonumber\\&+&\left.(g^{0})^{4}g_{1}^{0}e^{\big[4\kappa+\kappa_{1}+\big(\frac{1+2\alpha^{2}}{\alpha}
\big)\phi_{0}\big]r}dr^{2} \right. \label{23}.\ea Besides, the
physical BD field is given by
\be{\tilde{\phi}=\frac{1}{\mathcal{G}}e^{-2\alpha\phi_{0}r}}\label{CBD}.\ee

It is well known that the solution for the global cosmic string
leads to singularities in the spacetime, both in General
Relativity \cite{RUTHI} and in BD theory \cite{Linet}. This effect
is, in part, a characteristic of the dimension in question. From
the equation (\ref{15}) it is easy to see that, if the dimension
is not six ($p\neq 3$) the first term of the right-hand side is
not zero. This term leads to a differential equation other than
(\ref{22}). For example, in the Boisseau and Linet work $(p=1)$
the singularity for a finite value of $r$ is present. This
singularity comes from a contribution of such term. Fortunately,
in the present case it is suppressed. However the equation
(\ref{23}) is not free of singularities. For example, the Ricci
scalar is given by
\be{R=\frac{\Big[5\phi_{0}^{2}(4\alpha^{2}-1)-\alpha^{2}\kappa(3\kappa+2\kappa_{1})-2\alpha\kappa_{1}\phi_{0}+8\phi_{0}^{2}
\Big]}{(g^{0})^{4}g^{0}_{1}\alpha}e^{-\tilde{\sigma}
r}}\label{scalarRicci},\ee which means that, if the coefficient
$\tilde{\sigma} \equiv
4\kappa+\kappa_{1}+\Big(\frac{1+2\alpha^{2}}{\alpha}\Big)\phi_{0}$
of $r$ in the exponential factor is positive, the spacetime have a
singularity when $r\rightarrow -\infty$. Of course, if the
coefficient is negative the same thing happens to $r\rightarrow
+\infty$. This cumbersome situation can be solved if we split the
radial direction in two points, say $r=-r_{0}$ and $r=r_{c}$,
placing a new brane at each end points \cite{KMP}.

We will, however, take another direction by imposing an additional
constraint in the integration constants in order to have some
specific cases of analysis.

\section{Some specific examples}

The line element found in equation (\ref{23}) is very general
outside the string core. Potentially it can lead to a very
interesting variation of the current braneworld models. In this
section, we study some examples by imposing constraints in the
integration constants. Our hope is that they would serve as good
start points to more realistic models using global cosmic strings
in scalar-tensorial gravity theories.

As one can imagine, the choice of such constraints should be done
carefully. Let us analyze two examples where this choice can be
problematic. The first restriction one can, eventually, try is to
impose  $\tilde{\sigma}=0$. This case appears very interesting
since it leads to a constant curvature scalar, i. e., a maximally
symmetric spacetime. Apart of this, it is easy to note from
(\ref{23}) that such an imposition provides ``naturally'' an
exponential warp factor. However, this constraint is not possible
to be implemented. In fact, with the help of the equation
(\ref{u1}) one can relate $\kappa$ and $\kappa_{1}$ by
\ba(20\alpha^{2}+45-12\alpha^{4})\kappa^{2}+(24\alpha^{2}+30-8\alpha^{4})\kappa_{1}\kappa+(6\alpha^{2}+5)\kappa_{1}^{2}=0.\label{u3}\ea
Now, we can restrict the values of the constants $\kappa$ and
$\kappa_{1}$ by looking at the experimental bound of the
Brans-Dicke parameter. The current lower bound of the parameter
$w$ set it by $w\sim 10^{4}$ \cite{exper}. Hence, the most
important terms are proportional to $\alpha^{2}$ and there is not
a solution for $\kappa$ and $\kappa_{1}$ $\in \mathbb{R}$. The
same problem appears, for instance, if  the second term of the
right-hand side of equation (\ref{u1}) vanishes.

Let us remark an important assumption here. The bound to the
Brans-Dicke parameter is achieved from Solar System experiments.
Here, we are assuming that it is unchanged in a braneworld
scenario. In other words, we are assuming that the status of the
Brans-Dicke theory is also valid in the bulk. It is a strong
assumption. A complete scenario must take into account the
solution inside the brane, where this assumption is largely valid,
and then extend the result to the metric (\ref{23}) by some
junction conditions. In what follows, we still work with the
hypothesis that we can extrapolate the bound of $w$ to the brane
surface\footnote{We currently understand a point
$(x^{\mu},\theta,r=-r_{b})$, as belonging to the brane
``surface''.} $r=-r_{b}$ and to the bulk.

Since it is not possible to work with the hypothesis that
$\tilde{\sigma}=0$, let us take another one, this time with a more
direct relation between $\kappa$ and $\kappa_{1}$. It can be done
by \ba 3\kappa+2\kappa_{1}=0 .\label{ult1}\ea In this case, two
solutions are possible. From equation (\ref{u1}) one finds
\ba\phi_{0}^{\pm}=\frac{-3\kappa(1\pm
1)}{2\alpha(2+3/\alpha^2)}.\label{ult2}\ea Calling the exponential
factors in equation (\ref{23}) of \ba A\equiv
\kappa+\Big(\frac{1+2\alpha^2}{\alpha}\Big)\phi_{0},\label{ult3}\ea
\ba B\equiv
\kappa_{1}+\Big(\frac{-3+2\alpha^2}{\alpha}\Big)\phi_{0},\label{ult4}\ea
and \ba C\equiv
4\kappa+\kappa_{1}+\Big(\frac{1+2\alpha^2}{\alpha}\Big)\phi_{0},\label{ult5}\ea
we have, in the light of equations (\ref{ult1}) and (\ref{ult2}),
the following cases \ba A^{\pm}=\kappa\Bigg[1-\frac{3(1\pm
1)}{2}\Bigg(\frac{1+2\alpha^2}{3+2\alpha^2}\Bigg)\Bigg],\label{ult6}
\ea

\ba B^{\pm}=\frac{-3\kappa}{2}\Bigg[1+(1\pm
1)\Bigg(\frac{-3+2\alpha^2}{3+2\alpha^2}\Bigg)\Bigg],\label{ult7}
\ea and

\ba C^{\pm}=\frac{\kappa}{2}\Bigg[5-3(1\pm
1)\Bigg(\frac{1+2\alpha^2}{3+2\alpha^2}\Bigg)\Bigg].\label{ult8}
\ea The scalar curvature  has also a split into ``plus'' and
``minus'' solutions given by \ba R=\frac{9\kappa^{2}(1\pm
1)e^{-Cr}}{2(g^{0})^{4}(g_{1}^{0})(2\alpha^{2}+3)}\Bigg[\frac{(1\pm
1)}{2}[8+5(4\alpha^{2}-1)]-1\Bigg].\label{ult9}\ea

Now let us consider these two cases separately. The simplest one
is given by $\phi_{0}^{-}$. In that case we have $A=\kappa$,
$B=-3\kappa/2$, $C=5\kappa/2$, $R=0$ and $\phi_{0}=0$ (the
physical Brans-Dicke field is ``unplugged''). Hence, the metric is
simply given by \ba\tilde{ds^{2}}=e^{\kappa
r}g^{0}\eta_{\mu\nu}dx^{\mu}dx^{\nu}+g_{1}^{0}e^{\frac{-3\kappa
r}{2}}d\theta^{2}+(g^{0})^{4}g_{1}^{0}e^{\frac{5\kappa}{2}}dr^{2}.\label{ult10}\ea
In order to give the precise role of the warp factor we change
again the coordinates. So, by the relation \ba
d\rho^{2}=\bar{C_{1}}e^{Cr}dr^{2},\label{ult11}\ea where
$\bar{C_{1}}=(g^{0})^{4}(g_{1}^{0})$, we have \ba e^{\xi
r}=\Bigg(\frac{C\rho}{2\bar{C_{1}}^{1/2}}\Bigg)^{2\xi/C},\label{ult12}
\ea where $\xi$ is some constant. Note that the new coordinate
system runs in the range $\rho_{b}<\rho<+\infty$ where $\rho_{b}$
labels the brane surface. Now, the line element (\ref{ult10})
reads (absorbing $g^{0}$ and $g_{1}^{0}$ factors) \ba
\tilde{ds^{2}}=\Big(\frac{5\kappa
\rho}{4}\Big)^{4/5}\eta_{\mu\nu}dx^{\mu}dx^{\nu}+\Big(\frac{5\kappa
\rho}{4}\Big)^{-6/5}d\theta^{2}+d\rho^{2}.\label{ult13}\ea

The first thing to note from the above equation is the profile of
the line element. It is not well defined in the limit of zero
width $\rho_{b}\rightarrow 0$. Besides, the power law warp factor
shows a problematic behavior as $\rho \rightarrow +\infty$. We
emphasize that, apart of the fact that $R=0$ (and
$R_{ab}R^{ab}=0$, $a,b=0,...,5$), we expected a naked singularity
in the zero width limit since $R_{abcd}R^{abcd}\propto 1/\rho^{4}$
(and also $C_{abcd}C^{abcd}\propto 1/\rho^{4}$). To finalize the
analysis of this case, let us study the metric (\ref{ult13}) on
the brane surface $\rho=\rho_{b}$.

Note that the warp factor of $d\theta^{2}$ in (\ref{ult13}) is not
zero when $\rho=\rho_{b}$, the brane surface. So, keeping the
braneworld spirit in mind, we are forced to conclude that the
system consists of an extra transverse dimension and a 4-brane
with a compact curled extra dimension on-brane \cite{B6D}. This
means that the system presents an hybrid
compactification\footnote{On the other hand, if the warp factor of
the $\theta$ dimension presents an infinity number of zeros, it
will be possible to place infinity branes at each of these points
\cite{B6D} in the extra dimensional space. Note that, according to
this interpretation, questions about the behavior of the $\theta$
dimension outside the branes are meaningless. In other words, the
fact that the warp factor of $d\theta^{2}$ is able to take any
value in the range of $\rho$ is just important to fix the size of
this on-brane dimension.}. The spacetime described by
(\ref{ult13}) contains, then, a 4-brane with
$\mathbb{R}^{(3,1)}\times S^{1}$ topology placed in $\rho=0$ but,
about this, we do not have any information of the internal
structure. It is interesting the opposite behavior of the
exponential terms in (\ref{ult13}): for instance, as the warp
factor decreases with $\rho$, the size of the curled dimension
increases, and vice-versa. This behavior is also present in other
models in the Brans-Dicke gravity framework \cite{B6D}.

The size of the curled on-brane extra dimension should not
contradict the experimental bound \cite{experII}, say $M$. So,
from (\ref{ult13}) it is implemented by \ba \Big(\frac{5\kappa
\rho_{b}}{4}\Big)^{-6/5}<\frac{M}{2\pi},\label{u8}\ea which means
that
\ba\rho_{b}>\frac{4}{3\kappa}\Big(\frac{2\pi}{M}\Big)^{5/6}.\label{u9}\ea
The constraint (\ref{u9}) shows up the inviability of the solution
of the hierarchy problem, since a huge warp factor amplifies it.
In fact, the scale of the gravity interaction on the brane, say
$M_{6}$, is given by \ba
M_{6}^{4}=\frac{M_{Pl}^{2}}{2\pi}\Big(\frac{4}{3\kappa}\Big)^{6/5}\rho_{b}^{6/5},\label{enc1}\ea
where $M_{Pl}$ is the four dimensional Planck mass. Hence, taking
into account the relation (\ref{u9}), one sees that
$M_{6}>>M_{weak}$ and the hierarchy problem persists.

This type of behavior is also presented in other models \cite{UE}.
Of course, it is not a desirable characteristic for braneworld
models. However, we should emphasize that it is not a final word
since it depends explicitly of our previous choice of $\kappa_{1}$
and $\kappa_{2}$. Besides, we do not have any information about
the brane structure and it is possible that the shape of the warp
factor changes in the regime $\rho<\rho_{b}$.

Let us study one another possibility, $\phi_{0}^{+}$. As we will
see, this solution shows some interplay with the previous case in
the sense that the main conclusions are basically the same, apart
of a more interesting warp factor. In the case $\phi_{0}^{+}$ we
have in the old coordinate system
$A=\kappa\Bigg[1-3\Big(\frac{1+2\alpha^2}{3+2\alpha^2}\Big)
\Bigg]$,
$B=\frac{-3\kappa}{2}\Bigg[1+2\Big(\frac{-3+2\alpha^2}{3+2\alpha^2}\Big)
\Bigg]$,
$C=\frac{\kappa}{2}\Bigg[5-6\Big(\frac{1+2\alpha^2}{3+2\alpha^2}\Big)
\Bigg]$,
$R=\frac{9\kappa^{2}(20\alpha^{2}+2)}{(g^{0})^{4}(g_{1}^{0})(2\alpha^2+3)}e^{\frac{-\kappa(9-2\alpha^2)r}{2(3+2\alpha^2)}}$
and $\phi_{0}=\frac{-3\kappa \alpha}{3+2\alpha^2}$. Now, with the
help of equations (\ref{ult11}) and (\ref{ult12}) we arrive into
the following line element (absorbing the $g^{0}$, $g_{1}^{0}$,
$\kappa$ factors and some coefficients proportional to the
Brans-Dicke parameter) \ba
\tilde{ds^2}=\rho^{\big(\frac{-16\alpha^2}{9-2\alpha^2}\big)}\eta_{\mu\nu}dx^{\mu}dx^{\nu}+
\rho^{\big(\frac{18(1-2\alpha^2)}{9-2\alpha^2}\big)}d\theta^2+d\rho^2.
\label{ult14}\ea

Note that the relation between the warp factor and the $d\theta$
coefficient as $\rho$ changes is different from what we found in
the previous case. These two terms still have opposite behavior,
but now the exponent of the warp factor is negative, while the
exponential factor of the $d\theta$ coefficient is
positive\footnote{Of course, here we are again assuming the
possibility of extrapolation of the Brans-Dicke parameter data
obtained from solar system experiments.}. The scalar curvature
depends of $\rho$ as $R\propto 1/\rho^2$. So, we again expect a
naked singularity in the regime  $\rho_{b}\rightarrow 0$.

By the same reason we discussed in the previous case, here the
final scenario in composed by a $\mathbb{R}^{(3,1)}\times S^{1}$
brane and a transverse non-compact dimension. The solution on the
brane shows a more interesting case in which concerns to the
hierarchy problem. Restricting the size of the on-brane extra
dimension we have \ba
\rho_{b}<\Big(\frac{M}{2\pi}\Big)^{(9-2\alpha^2)/9(1-2\alpha^2)}\label{ult15}.\ea
However, this very small value for $\rho_{b}$ can be, at least in
part, compensated by the minuteness of the Brans-Dicke parameter
and the warp factor appears not so huge.

To finalize this Section we stress that the physical Brans-Dicke
field obeys the following power law $\tilde{\phi}\sim
\rho^{24\alpha^2/(9-2\alpha^2)}$ which, in the light of equation
(\ref{ult15}), is very weak on the brane. We shall  comment about
the main results in the next Section.

\section{Final remarks and outlook}

To start with, we solved the EBD gravitational equations using a
global cosmic string as a source in six dimensions. The topology
of the brane is given by $\mathbb{R}^{(3,1)}\times S^{1}$, i. e.,
an hybrid compactification.  Of course, the topology of the bulk
is not a direct product of $\mathbb{R}^{(3,1)}\times S^{1}$ and
some extra dimension geometry since the dimensions are mixed up
\cite{Sun}. An important issue would be to extend this topology to
encode an orbifold in both extra directions. In other words, to
replace the on brane $S^{1}$ by $S^{1}/\mathbb{Z}_{2}$ and try to
incorporate this orbifold in the non-compact dimension too. In
such case the matching conditions are well established and one can
analyze some interesting properties of physical systems \cite{ROY}
such as black-holes on the brane.

Going forward, in order to extract some physical conclusions, we
particularized a specific range for the integration constants in
order to study some specific concrete examples. This analysis
shows up an amplification of the hierarchy problem by a huge warp
factor in one of the examples. In the second example this problem
is partially circumvented by the Brans-Dicke parameter appearing
in the power law warp factor.

One can study the linearized tensor fluctuations in order to
localize gravity in the model. Keeping the classical approach to
braneworld models, such a fluctuation is very important as well as
the linearized tensor fluctuation is the analysis of the
propagation of scalars \cite{CEG,UE} in the background
established. We will not do it here since the model is not
complete in the sense that we made some specific choices for the
constants. As we said before, a complete solution should
contemplate the line element in the region $\rho<\rho_{b}$ (or
$r<-r_{b}$), i. e., inside the brane. Normally, there are some
standard approaches for the extension of solutions in the brane
surface to the brane interior: it can be done, for example, by the
analysis of the scalar propagator in the limit
$\rho_{b}\rightarrow 0$ \cite{CEG}, or by some trick as the
imposition of the  $\mathbb{Z}_{2}$ symmetry at the brane surface
\cite{UE}. Unfortunately, these two approaches are not possible in
this model since it is based in a real topological defect, the
global cosmic string, within the Brans-Dicke theory. In this vein,
perhaps a more insightful approach is to find a solution splitting
the spacetime into two regions (inside and outside the string) and
finding the appropriated matching conditions, in a similar way of
what it was done in the reference \cite{ins1}. Besides, the moduli
of the extra on-brane dimension can be stabilized by such junction
conditions. This type of consideration should be taken seriously
in the continuation of this research line.

To conclude, we should emphasize that the study of
phenomenological implications of extra dimensions is a very
interesting and promising field of research. In particular there
is a great amount of work dealing with the possibility of some
dependence of the standard model fields on extra dimensions, the
curled and also the noncompact ones \cite{fenomenologia}. In the
case of the dependence of the curled on brane dimension, for
example, it is well known that the Kaluza-Klein tower, or
specifically the lightest particle of such spectrum, can supply a
good candidate to dark matter. Going into this direction, one can
make an important and fundamental link between extra dimensional
models and high energy fenomenology.

\section*{Acknowledgments}

The authors benefited from useful discussions with Profs. Andrey
Bytsenko, Jos\'e Helay\"el-Neto, Rold\~ao da Rocha, Marcelo
Azevedo Dias and Hiroshi de Sandes Kimura. We also thanks a
referee for so many important remarks on the manuscript as well as
for bring our attention about important literature on this
subject. J. M. Hoff da Silva thanks CAPES-Brazil for financial
support. M. E. X. Guimar\~aes and M. C. B. Abdalla acknowledge
CNPq for support.

\end{document}